\documentclass[useAMS]{mn2e}
\usepackage{graphicx}
\usepackage{epsfig}

\def\ltsima{$\; \buildrel < \over \sim \;$}
\def\lsim{\lower.5ex\hbox{\ltsima}}
\def\gtsima{$\; \buildrel > \over \sim \;$}
\def\gsim{\lower.5ex\hbox{\gtsima}}

\newcommand{\be}{\begin{equation}}
\newcommand{\en}{\end{equation}}

\begin{document}
\title[Effective column densities in GRB X--ray spectra]{Effective absorbing column density in the gamma--ray burst afterglow X--ray spectra}

\author[Campana et al.]{S. Campana$^{1,}$\thanks{E-mail: sergio.campana@brera.inaf.it}, 
M. G. Bernardini${^1}$, V. Braito$^1$,  G. Cusumano$^2$, P. D'Avanzo${^1}$, \newauthor
V. D'Elia$^{3,4}$,  G. Ghirlanda${^1}$, G. Ghisellini${^1}$, A. Melandri$^1$,   R. Salvaterra${^5}$, \newauthor
G. Tagliaferri${^1}$, S. D. Vergani$^{6,1}$\\ 
$^1$ INAF - Osservatorio astronomico di Brera, Via Bianchi 46, I--23807, Merate (LC), Italy\\
$^2$ INAF - Istituto di Astrofisica Spaziale e Fisica Cosmica di Palermo, Via U. La Malfa 153, I-90146, Palermo, Italy\\
$^3$ ASI - Science Data Center, Via Galileo Galilei, I--00044, Frascati (Roma), Italy\\
$^4$ INAF - Osservatorio Astronomico di Roma, Via Frascati 33, I--00040, Monteporzio Catone (Roma), Italy\\
$^5$ INAF - IASF Milano, Via Bassini 15, I--20133 Milano, Italy\\
$^6$ GEPI, Observatoire de Paris, CNRS, Univ. Paris Diderot, 5 place Jules Janssen, 92195, Meudon, France
}

\maketitle

\begin{abstract}
We investigate the scaling relation between the observed amount of
absorption in the X--ray spectra of Gamma Ray Burst (GRB) afterglows and the 
absorber redshift.
Through dedicated numerical simulations of an ideal instrument, we
establish that this dependence has a power law shape with index 2.4.
However, for real instruments, this value depends on their
low energy cut-off, spectral resolution and on the detector spectral response
in general. We thus provide appropriate scaling laws for specific instruments.
Finally, we discuss the possibility to measure the absorber redshift from X--ray data
alone. We find that $10^{5-6}$ counts in the 0.3--10 keV band are needed
to constrain the redshift with $10\%$ accuracy. As a test case we discuss
the XMM-Newton observation of GRB 090618 at $z=0.54$. We are able to
recover the correct redshift of this burst with the expected accuracy.
\end{abstract}


\section{Introduction}

Radiation emitted by distant objects gets absorbed by matter along the line of sight. 
For extragalactic objects this absorption process occurs within our Galaxy, 
in the intergalactic medium (IGM) and within the host galaxy.

Long gamma--ray bursts (GRBs) are associated with the death of massive stars in distant galaxies.
Their gamma--ray radiation is not absorbed by intervening matter but this occurs to their afterglow
lower energy radiation at X--ray and optical frequencies.
X--ray absorption is due to the innermost shell transitions in metals
irrespective of their form (gas, dust, etc.). X--ray absorption in the GRB afterglow spectra 
shows up as a characteristic bending of the power law at low energies ($\lsim 1$ keV) depending
on the amount of material along the line of sight.
 
X--ray afterglow radiation gets absorbed within the host galaxy, within our Galaxy and possibly along these two galaxies 
by diffuse matter (see Eitan \& Behar 2013; Behar et al. 2011) and/or by intervening collapsed systems 
(Campana et al. 2010, 2012; Wang 2013).
The extragalactic absorption is inherently shifted in energy due to the cosmological redshift.
The same amount of absorption, placing the absorber at increasing redshifts, produces an increasingly
small effect on the observed X--ray spectrum, due to the shift in energy.
The scaling of a local absorption at a given redshift $N_H(z)$ with the redshift has been scaled 
as $(1+z)^a$ with $a$ ranging in the $2.4-2.6$ range (e.g. Galama \& Wijers 2001; Stratta et al. 2004; Campana et al. 2006a; 
Watson et al. 2007; Behar et al. 2011; Watson \& Jakobsson 2012; Starling et al. 2013).
For the case of GRB 090423 (Salvaterra et al. 2009;  Tanvir et al. 2009) at $z=8.2$ the discrepancy due to
the different scalings with redshift (i.e. values of $a$) can be as high as $\sim 60\%$.

\begin{figure*}
\begin{center}
\begin{tabular}{cc}
\includegraphics[width=5.5cm,angle=-90]{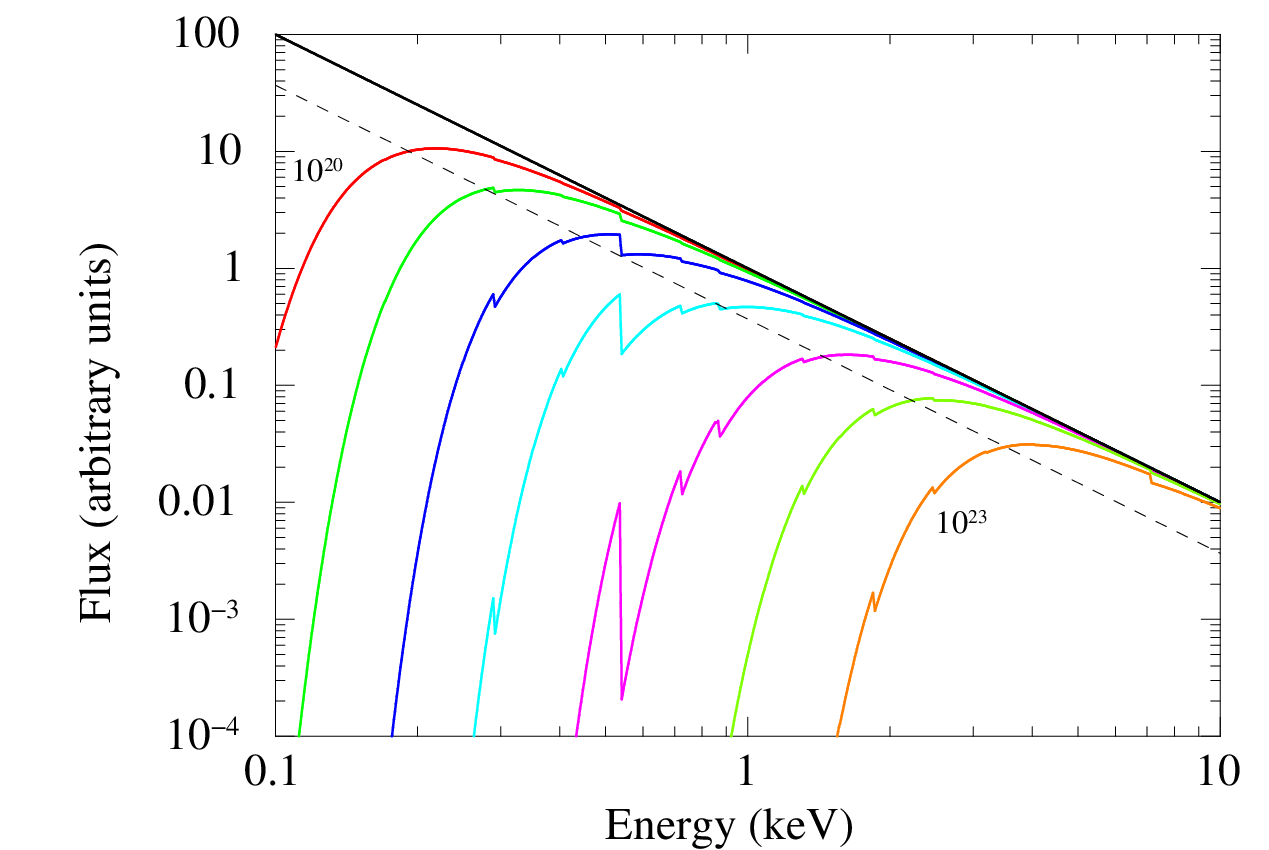} &
\includegraphics[width=5.5cm,angle=-90]{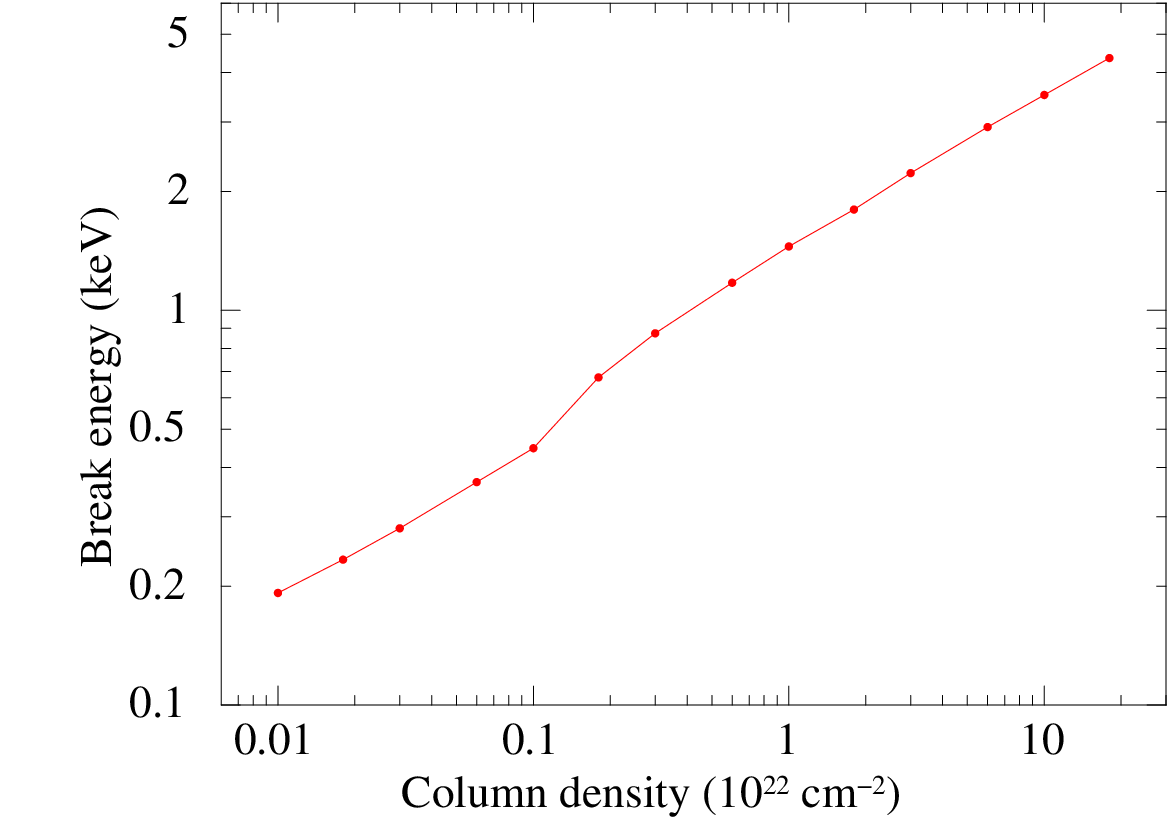} \\
\end{tabular}
\end{center}
\caption{Left: Simulated power law spectra with increasing column density Galactic absorbers. From top to bottom: 
0, $10^{20}$, $3\times10^{20}$, $10^{21}$, $3\times10^{21}$, $10^{22}$, $3\times10^{22}$, $10^{23}$ cm$^{-2}$, respectively.
The dashed line marks the $e-$folding energy $E_e$.
Right: Relation between the absorbing column density $N_H$ and the break $e-$folding energy $E_e$. A higher sampling has been adopted 
to better constrain the power law shape.}
\label{break}
\vskip -0.1truecm
\end{figure*}

Here we investigate the scaling of the intrinsic column density $N_H(z)$ with $z$ depending on its value, on the Galactic 
column density and on the instrument energy band. We derive a universal scaling. This scaling relation is proved to differ slightly 
depending on the mission/instrument considered (Section 2).
In principle, with a very large number of photons one can disentangle the contribution of the Galactic absorption $N_H^{\rm Gal}$ 
from the intrinsic absorption at a given redshift and derive from the fit the redshift of the absorber, if the spectral shape
is known as in the case of GRBs (i.e. a power law). In Section 3 we use this scaling to set the minimum number of counts needed to 
derive the GRB's redshift for several values of the intrinsic column density and 
we work out an example of this technique, based on XMM-Newton data of GRB 090618.
Conclusions are reported in Section 4.

\section{Intrinsic column density scaling with redshift}

As a first step we study the influence of an increasing column density with energy.
Absorption cross sections was firstly computed for astronomical purposes by Morrison \& McCammon (1983).
For a given value of the equivalent column density one can associate an energy corresponding to, e.g., 
$1/e$ decrease of the source intensity, $E_e$.
We carried out simulations using the spectral package XSPEC (v. 12.8.0.m; Arnaud 1996). 
We modelled the absorber using the {\tt TBABS} model within XSPEC (Wilms, Allen \&McCray  2000) 
and performed simulations with the {\tt wilm} solar abundance pattern and the {\tt vern} photoelectric absorption cross sections
(Verner et al. 1996).
We simulated a model made by a power law with photon index $\Gamma=2$ and for a range of values of the 
equivalent column density $N_H$ we computed the corresponding $e-$folding energy $E_e$.
Results were shown in Fig. \ref{break}. Similar  $e-$folding energies were obtained with different power law spectral slopes.
There are two different slopes in the $E_e-N_H$ dependence due to the Oxygen edge intensity (at 0.543 keV, the most prominent edge 
in the X--ray spectrum). For low column density values the edge produces a decrease lower than $1/e$, whereas for high column densities 
the $1/e$ drop occurs across the Oxygen edge. For energies lower than $\sim 0.1$ keV, we have 
$N_H\propto E_e^{2.67}$ and for energies above $\sim 0.2$ keV $N_H\propto E_e^{2.61}$. However, the overall dependence is $\sim 2.3$,
which is not close to 2.6.

\begin{figure}
\begin{center}
\includegraphics[width=5.5cm,angle=-90]{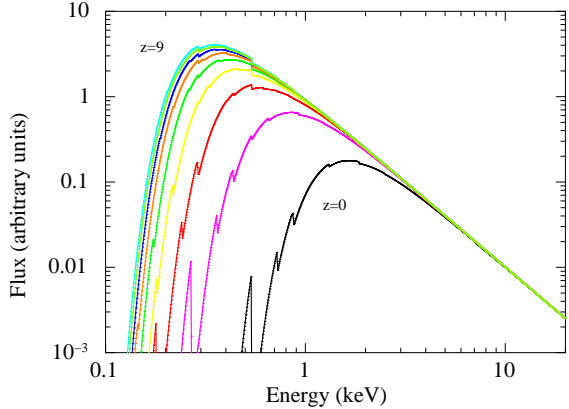} 
\end{center}
\caption{Simulated power law spectra ($\Gamma=2$) with different absorption. A Galactic column density of $N_H^{\rm Gal}=3\times 10^{20}$ 
cm$^{-2}$ is common to all the spectra. In addition a large column density of $N_H(z)=10^{22}$ cm$^{-2}$ is added for different values of the redshift $z$.
 From bottom to top the redshifts are range from $z=0$, to $z=9$ in step of one. }
\label{nhzz}
\vskip -0.1truecm
\end{figure}

We then investigated the dependence of the intrinsic column density $N_H(z)$ with the redshift,
with a scaling law of $(1+z)^a$. The effect of the redshift on the column density is to shift in energy the absorbed part,
resulting in a lower effective column density (see Fig. \ref{nhzz} and compare it to Fig. \ref{break}).  In order to do this we simulated a power law spectrum with a 
spectral slope $\Gamma=2$, using a diagonal redistribution matrix in the 0.01--20 keV energy range and flat effective area of 1 cm$^2$.
The GRB afterglow was assumed to be very bright (normalisation constant within XSPEC equal to 1000, for an exposure time of 1000 s).
This was done to estimate the column densities without redistribution biases and with a large number of counts.
As a first step we did not consider any Galactic absorption and we just included an intrinsic absorber $N_H(z)$. 
We took $N_H(z)=10^{23}$ cm$^{-2}$. We modelled the absorber with the {\tt ZTBABS} model within XSPEC (Wilms et al.  2000). 
The metallicity $Z$ was assumed to be solar. This is not crucial since, at first order, the metallicity contribution scales 
directly with the column density, so that $N_H/Z=const$.
We carried out simulations for different GRB redshifts in the 0--9 redshift range in step of 1.
Each simulated spectrum was then fit with a power law model and an absorber at redshift $z=0$, providing an effective column density $N_H(z=0)$.
Given the ``wrong redshift" of the column density absorber (i.e. we are fitting with a Galactic $z=0$ column density absorber something that, instead, 
is intrinsic to the host galaxy at a given redshift $z$), the fits were not statistically acceptable, but we can derive anyway a value for 
$N_H(z=0)$, together with an error based on $\Delta \chi^2=2.7$. Given the large simulated statistics the largest deviations occurred around
the absorption edges.
The values of  $N_H(z=0)$ are therefore linked to the intrinsic column density values by the law:
$$
N_H(z=0)=N_H(z)/(1+z)^a
$$

\begin{table}
\caption{Scaling laws of the effective absorbing column density depending on the low energy cut-off.}
\label{fitcutoff}
\begin{center}
\begin{tabular}{cc}
\hline
Low energy cut-off   &Scaling index    \\
(keV)                           & $a$                       \\
\hline
0.01                             &$2.40\pm0.01$\\
0.03                             &$2.40\pm0.01$ \\
0.1                               &$2.40\pm0.01$\\
0.3                               &$2.34\pm0.01$ \\
0.5                               &$2.39\pm0.01$ \\
0.7                               &$2.51\pm0.01$\\
1.0                               &$2.60\pm0.01$ \\
\hline
\end{tabular}
\end{center}
\end{table}

\begin{figure}
\begin{center}
\includegraphics[width=5.5cm,angle=-90]{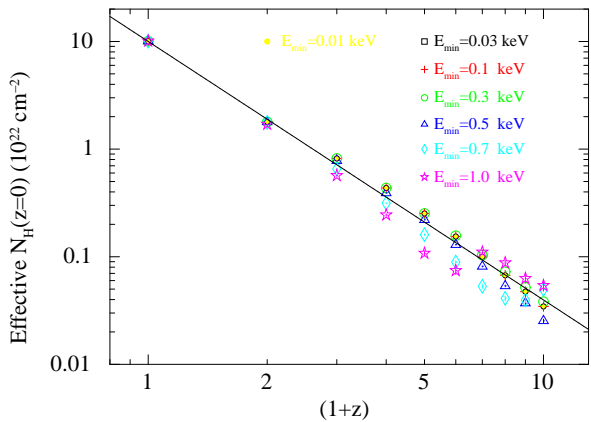}
\end{center}
\caption{Effective column density $N_H(z=0)$ versus redshift as evaluated through simulations 
of the X--ray spectrum of a GRB afterglow. In this case we varied the low-energy cut-off of the X--ray instrument.
Sharp changes in the slopes are due to the major edges (Oxygen and Iron) moving outside the energy band.}
\label{cutoff}
\vskip -0.1truecm
\end{figure}

We found that the scaling index is $a=2.4$ (see Table \ref{fitcutoff}, with small values of the instrument low energy cut off).
We first explored the dependence of the scaling index with the adopted power law photon index $\Gamma$.
The dependence on $\Gamma$ was mild: for $\Gamma=1$ we derived $a=2.39$ and for $\Gamma=3$ $a=2.44$.
We also explored the dependence of the scaling parameter $a$ on the low energy cut-off of an ideal X--ray instrument.
We progressively increased the low energy instrumental cut-off from 0.01 keV to 1.0 keV.
Results were reported in Table \ref{fitcutoff} (see also Fig. \ref{cutoff}).
We can observe a progressive steepening of the scaling index from 2.4 to 2.6 when the cut-off is around 1 keV.
The shift in energy of the contribution of the intrinsic column density reaches a point where the 
curvature of the spectrum (i.e. the $e-$folding energy $E_e$) goes below the adopted low energy cut-off. 
At this point the fit tends to underestimate the overall absorption resulting in a steeper dependence.

\begin{figure}
\begin{center}
\includegraphics[width=5.5cm,angle=-90]{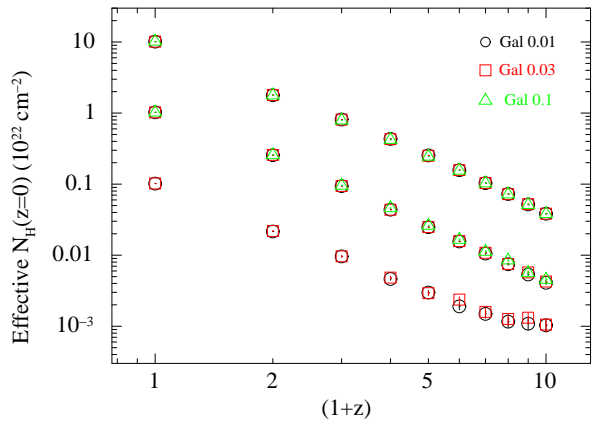}
\end{center}
\caption{Effective column density $N_H(z=0)$ versus redshift as evaluated through simulations 
of the X--ray spectrum of a GRB afterglow. Three different sets of Galactic column densities were used
(0.01, 0.03 0.1 $\times 10^{22}$ cm$^{-2}$, marked with circles, squares and triangles, respectively),
for three different initial values of the intrinsic column density $N_H(z)$, 0.1, 1 and 10 $\times 10^{22}$ cm$^{-2}$. }
\label{simule}
\vskip -0.1truecm
\end{figure}

We then turned on the Galactic column density. We investigated three cases with increasing Galactic column densities:
$N_H^{\rm Gal}=10^{20}$, $3\times 10^{20}$ and $10^{21}$ cm$^{-2}$. These are typical values for line of sights 
out of the Galactic plane.
We run the simulations with a sample of intrinsic column densities over the 0--9 redshift range as above (see Table \ref{fitsimule} and
Fig. \ref{simule}). In this case we realistically limit the low energy range of the instrument to 0.3 keV.
We derived flatter indexes and we were able to almost recover the 
original scaling law $a=2.34$ (appropriate for a 0.3 keV low energy cut-off) only for very high intrinsic column densities, 
where the contrast with the Galactic column density is the highest.
We could also compute the redshift, $\bar{z}$, at which the effective contribution of the intrinsic column density becomes comparable 
to the Galactic contribution (i.e. $N_H(z=0)\sim N_H^{\rm Gal}$).
These redshifts were also reported in Table \ref{fitsimule}.
These simulations were carried out assuming a perfect knowledge of the Galactic column density. 
Despite the intrinsic uncertainties in the radio maps (usually assumed of about $\sim 10\%$; Kalberla et al. 2005; Dickey \& Lockman 1990),
recently other prescriptions were put forward to estimate the Galactic column density based on dust maps (Watson 2011), or
including the contribution of molecular hydrogen (Willingale et al. 2013), possibly increasing the column density uncertainty.
The net effect of an uncertain Galactic column density is to increase the uncertainty on the intrinsic column density.
In particular, for values of the redshift equal or larger than $\bar{z}$ redshift, we usually derived upper limits on the intrinsic column density.
We explored the effect of a $30\%$ uncertainty on $N_H^{\rm Gal}$ on the highest intrinsic column densities reported in Table 
\ref{fitsimule}. The effect was to produce a mild flattening of the scaling index $a$ of $\sim 0.03$.

\begin{table}
\caption{Scaling laws of the effective absorbing column density.}
\label{fitsimule}
\begin{center}
\begin{tabular}{cccc}
\hline
Galactic $N_H^{\rm Gal}$&Intrinsic  $N_H(z)$       &Scaling index  & Redshift  \\
($10^{22}$ cm$^{-2}$)      &($10^{22}$ cm$^{-2}$)& $a^*$                  &  equality   ($\bar{z}$)    \\
\hline
0.01                                       & 0.1                                 &2.20 (2.24) & 1.6\\
0.01                                       & 1                                    &2.30 (2.28) &  5.8\\
0.01                                       & 10                                  &2.38 (2.38) & 16.8 \\
\hline
0.03                                       & 0.1                                 &2.19 (--)     & 0.7\\
0.03                                       & 1                                    &2.29 (2.22)& 3.3\\
0.03                                       & 10                                  &2.37 (2.37)& 10.2 \\
\hline
0.1                                         & 1                                    &2.23 (1.99)& 1.6\\
0.1                                         & 10                                  &2.35 (2.26)& 5.8 \\
\hline
\end{tabular}
\end{center}
\noindent $^*$ In parenthesis is the scaling index over the redshift range in which the effective column density is larger 
than the Galactic column density.
\end{table}

Finally we investigated the dependence of the scaling law on the spectral resolution (variable along the energy range) 
and on the detector response in general.
We simulated spectra as above (excluding the Galactic column density) adopting different response matrices taken from the WebSpec 
site\footnote{http://heasarc.gsfc.nasa.gov/webspec/webspec.html} for several on orbit instruments.
For completeness, we also included a low spectral resolution instrument (BeppoSAX LECS) and a high spectral 
resolution calorimeter (Athena).
We found that the scaling law is different for different instruments (see Table \ref{fitmission}). As a general rule, instruments with a high 
spectral resolution did not show differences with the simulations carried out with the diagonal response matrix (compare Tables
\ref{fitsimule} and \ref{fitmission}), being the low energy cut-off the only parameter changing the scaling law index $a$.
For instruments with a lower spectral resolution, which is also variable across the spectral energy range,
we found in some cases  differences with the diagonal response matrix simulations.
This indicates that also the spectral resolution can affect the scaling law and that,
when dealing with GRB spectra of one specific instrument, one should work out and use the appropriate scaling law
in a specific energy range.

If some intervening system(s) or diffuse system were located along the line of sight, they will alter the absorption modelling described above.
A localised system will be characterised by its own redshift and will imprint on the GRB spectrum a characteristic signature at
its own redshift. Given the present generation of X--ray instruments, it is very difficult to directly detect the imprint of an 
intervening system on the spectrum of a GRB. What can happen instead is that one does not know about the presence
of this intervening system at redshift $z_{\rm int}$. Therefore one is tempted to ascribe the full absorption pattern in excess of the Galactic value
to the host galaxy of the GRB at a redshift $z_{\rm GRB}$. In this way the intervening system column density is artificially increased
by a factor $(1+z_{\rm GRB})^a/(1+z_{\rm int})^a$, producing an increase of the column densities of GRBs along with the redshift
(e.g. Campana et al. 2010, 2012). The same effect occurs for a diffuse medium along the line of sight between our Galaxy and the 
GRB host galaxy (Behar et al. 2011).
It must be noted, however, that of all the well studied intervening systems observed so far (e.g. Sparre et al. 2013), 
none is able to account for the large observed discrepancy among X--ray and optically determined column densities (Campana et al. 2010;
Watson et al. 2007).

\begin{table}
\caption{Scaling laws of the effective absorbing column density for different X--ray instruments.}
\label{fitmission}
\begin{center}
\begin{tabular}{ccc}
\hline
Mission \&             & Low energy cut-off   &Scaling index   \\
Instrument             & (keV)                           & $a$                    \\
\hline
Swift-XRT-PC       & 0.3                              &$2.34\pm0.01$   \\
Chandra-ACIS-S  & 0.5                              &$2.46\pm0.03$ \\
Chandra-LETG-1 & 0.2                              &$2.42\pm0.03$\\
XMM-pn-thin         & 0.2                              &$2.35\pm0.01$ \\
Suzaku-BI              & 0.4                              &$2.38\pm0.01$\\
BeppoSAX-LECS & 0.1                             &$2.42\pm0.01$\\
Athena-Cal.           & 0.15                           &$2.39\pm0.01$\\
\hline
\end{tabular}
\end{center}
\end{table}

\section{Determining the redshift from GRB afterglow data}

The imprint of the intrinsic absorber on a GRB afterglow X--ray spectrum can,
at least in principle, lead to the determination of the absorber's redshift.
In fact, the effect of the absorbing column density is to bend the afterglow power law spectrum 
but also to leave an imprint of the absorption edges at characteristic energies.
The observed shape of the bend at low energies and the observed energy position of the edges depend of the absorber's redshift.
If the number of counts is sufficiently high one can recover the GRB redshift directly from the 
X--ray data, once the Galactic column density is known with good accuracy.

\begin{figure}
\begin{center}
\includegraphics[width=5.5cm,angle=-90]{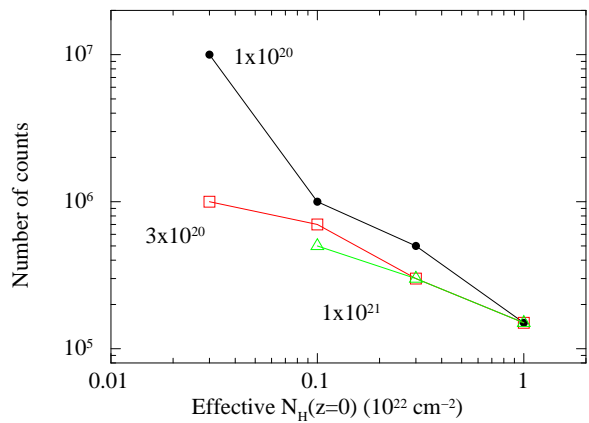}
\end{center}
\caption{Number of photons needed to determine the redshift of the absorber for different 
values of the effective column density $N_H(z=0)$. Three different Galactic column densities
have been considered: 0.01, 0.03 0.1 $\times 10^{22}$ cm$^{-2}$, marked with circles, squares and triangles.}
\label{fotoni}
\vskip -0.1truecm
\end{figure}

\begin{figure*}
\begin{center}
\begin{tabular}{cc}
\includegraphics[width=5.5cm,angle=-90]{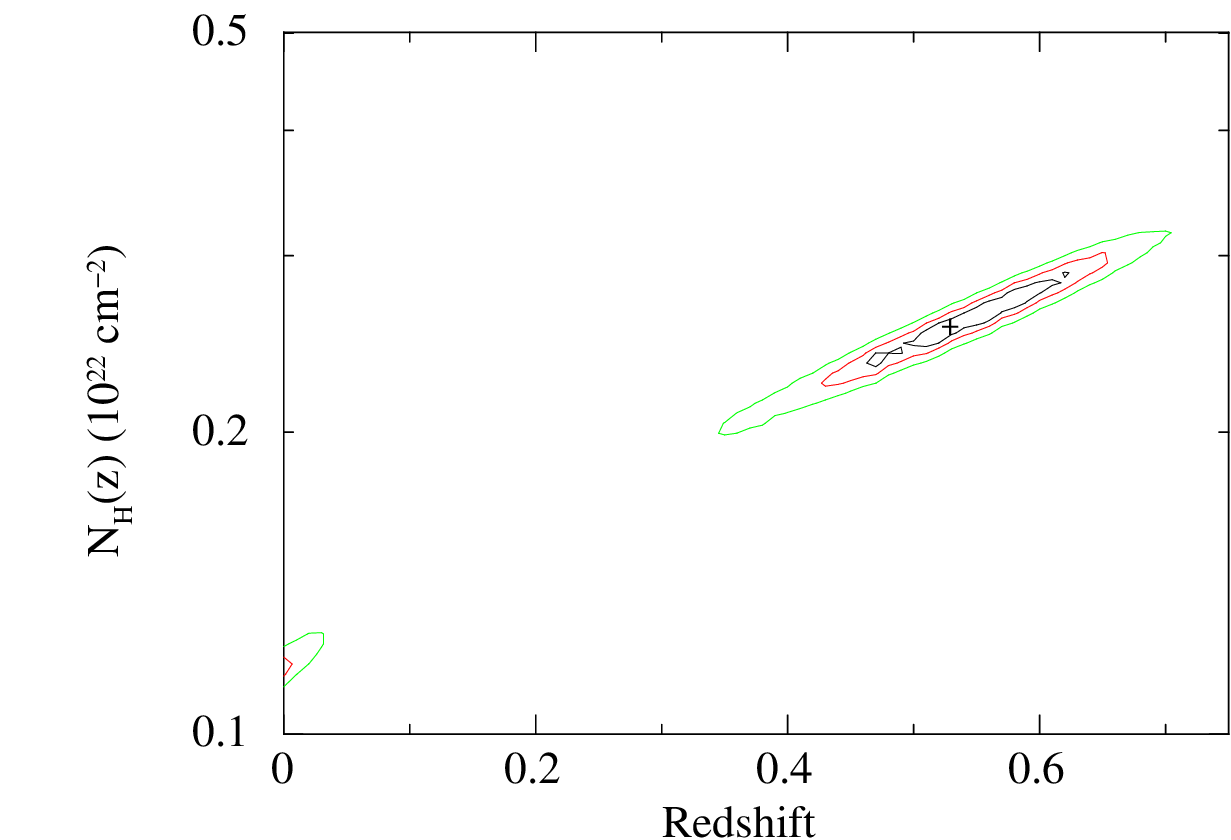} &
\includegraphics[width=5.5cm,angle=-90]{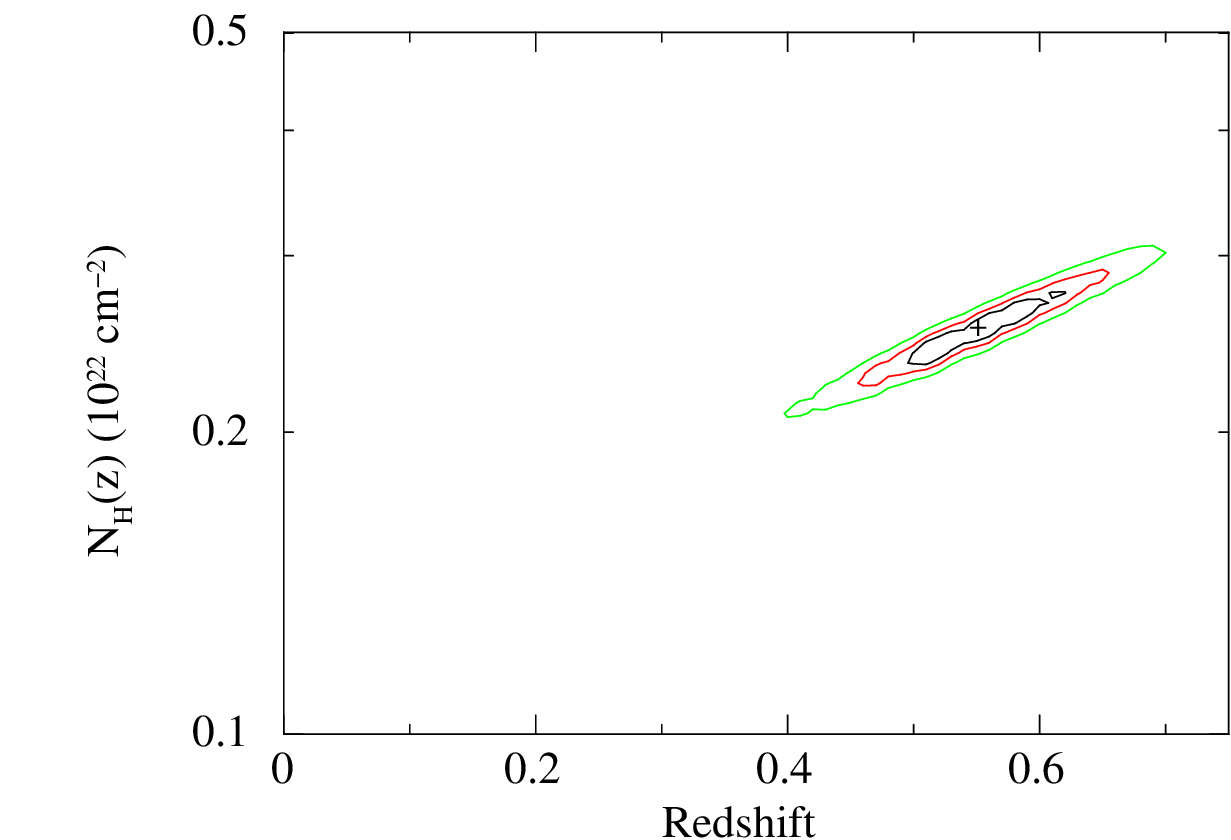} \\
\end{tabular}
\end{center}
\caption{Contour plots of the redshift versus intrinsic column density of GRB 090618 based on XMM-Newton data.
The left panel is a zoom of the MOS+pn data in the $z=0-1$ region (data were searched up to $z=8$). The right panel shows 
the effect of the addition of RGS instruments data, ruling out the zero redshift (low statistics) solution and improving the redshift solution.}
\label{nhz}
\vskip -0.1truecm
\end{figure*}

Given the scaling relations derived in the previous section, 
we can combine the intrinsic column density $N_H(z)$ and the redshift $z$ into an effective column density $N_H(z=0)$.
Clearly, the redshift determination will be easier for a larger intrinsic column density and/or a smaller redshift
(because as the redshift rises also the effective column density decreases).
Based on Swift XRT (Burrows et al. 2005) response matrices (we adopted the redistribution matrix {\tt swxpc0to12s6\_20010101v013.rmf} 
and the ancillary response file {\tt swxpc0to12s6\_20010101v013.arf}; Beardmore et al. 2013) we worked out the minimum 
number of counts needed to derive the redshift from the X--ray data alone with a$\lsim 10\%$ uncertainty.
We considered three Galactic column density values  logarithmically spaced ($N_H^{\rm Gal}=3\times 10^{20}$, $10^{21}$ and $3\times 10^{21}$ cm$^{-2}$).
For a given Galactic column density, we simulated a power law spectrum with $\Gamma=2$ for four 
effective column densities $N_H(z=0)$ at different redshifts ($3\times 10^{20}$, $10^{21}$, $3\times 10^{21}$ and $10^{22}$ cm$^{-2}$). 
Each effective column density was obtained by varying the redshift between 1, 2, 4 and 6 and computing the 
intrinsic column density $N_H(z)$ according to the scaling relation above, in order to have the 
same effective column density $N_H(z=0)$.
For each value of the effective column density and of the Galactic value we fitted the spectra of the 
four realisations ($z=1,\  2, \ 4, 6$), which at first order provide the same effects on the overall X--ray spectrum, 
and took the one with the highest number of counts needed to constrain the redshift with an error $\Delta z/z\lsim 10\%$ 
($90\%$ confidence level).
The results of these simulations were shown in Fig. \ref{fotoni}. It is readily apparent that we need a very large number of counts.
When the effective column density was close to the (known) Galactic value the number is $\gsim 10^6$ counts.
As long as the effective column density increased over the Galactic value it becomes easier and easier to 
derive the redshift with $\sim 10^5$ counts in the best cases investigated.

\subsection{A real case: XMM-Newton observations of GRB 090618}

As a test case we consider the XMM-Newton observations of GRB 090618.
This is one of the brightest GRBs observed with XMM-Newton. Indeed the brightest GRB observed by XMM-Newton 
was GRB 060729 (Grupe et al. 2007) but its effective column density is low.
GRB 090618 provides the best case in having a high effective column density and a large number of photons
(Campana et al. in preparation).
GRB 090618 was discovered by Swift and observed by XMM-Newton within 5.3 hr from the  
trigger. The GRB redshift is $z=0.54$ (Cenko et al. 2009). The Galactic column density is $5.7\times 10^{20}$ cm$^{-2}$ (Kalberla et al. 2005).
MOS (0.3--10 keV) and pn (0.2--10 keV) data provide $\sim 135,000$ counts. A comparable, or in some case even larger, number of counts are 
provided by very bright GRBs promptly observed by the Swift-XRT. However, in these cases spectral variations are often observed (e.g. the softening 
accompanying the steep decay phase, Tagliaferri et al. 2005, or the evolving black body component in some nearby GRBs, Campana et al. 2006b).
Spectral changes were proved to alter the column density estimate (Butler \& Kocevski 2007), making unfeasible our procedure.
XMM-Newton data analysis was throughly described in Campana et al. (2011a). 

Fitting the spectrum with an absorber at $z=0$ we derived $N_H(z=0)=(1.17\pm0.04)\times 10^{21}$ cm$^{-2}$ (in addition to the Galactic 
value), with a reduced $\chi^2_{\rm red}=1.020$ for 1391 degrees of freedom (dof, $30\%$ null hypothesis probability, nhp).
Based on data in Fig. \ref{fotoni} we would need $\sim 400,000$ counts to derive the redshift with a $10\%$
accuracy ($90\%$ confidence level).
Leaving free the redshift the fit improved to $\chi^2_{\rm red}=1.018$ (1390 dof and $31\%$ nhp), obtaining the 
correct redshift of $z=0.55^{+0.08}_{-0.10}$ ($90\%$ confidence level). This is a $\sim 18\%$ accuracy redshift determination. 
An F-test indicates a mild improvement in the fit with a chance probability of $8\%$. 
We investigated thoroughly the $N_H(z)-z$ plane and found a low significance ($\sim 3\,\sigma$) zero-redshift solution (see Fig. \ref{nhz}). 
The zero-redshift solution implied $z<0.03$ at a $3\,\sigma$ confidence level. There are just two GRBs within this redshift range
(GRB 980425, Galama et al. 1998 and GRB 060218 Campana et al. 2006b) and for both of them a host galaxy is clearly visible, as 
well as a bright accompanying supernova. For these reasons the zero-redshift can be easily discarded.
We noted that the addition of 0.45--1.8 keV RGSs data (2,924 and 3,482 counts for the RGS1 and RGS2, respectively), remove this (small) 
degeneracy, thanks to the higher spectral resolution, and largely improved the non-zero redshift significance
(see Fig. \ref{nhz}). The RGS data alone constrained the redshift to the $z=0.1-5.4$ interval. The inclusion of the RGS data led 
to a redshift $z=0.55^{+0.08}_{-0.06}$ and to an improvement over the zero redshift solution of $\sim 3\,\sigma$ (based on an F-test).

\section{Conclusions}

We investigated the dependence on the redshift $z$ of the effective absorption of X--ray photons produced by a localised
system at a given redshift. Such intrinsic absorption is commonly observed in the X--ray spectra of GRB afterglows.
We worked out the dependence of the effective (i.e. observed frame) column density $N_H(z=0)$ on $(1+z)$ 
based on spectral simulations. This dependence has a power law shape with index $a=2.40$. 
This settles an issue since values in the 2.4--2.6 range were used in the past.

We investigated the dependence of the scaling law on the low energy cut-off of the X--ray instrument finding 
a small decrease ($a=2.34$) when the low-energy cut-off is $\sim 0.3$ keV, to increase up to $a\sim 2.6$ 
for cut-off energies as large as 1 keV. 
We tested the scaling relation for several instruments finding that, in addition to the low energy cut-off, also the 
(variable) spectral resolution has an impact on the value of $a$. We proved that an instrument-specific relation should 
be used to rescale the intrinsic column densities of GRBs to zero redshift (see Table \ref{fitmission}). 
Grating or calorimeters provided values of $a$ closer to the diagonal matrix solutions.

We also tested, in the case of the Swift XRT instrument, the minimum number of photons needed to evaluate 
directly from X--ray data the redshift of the GRB with a $\sim 10\%$ accuracy, in the hypothesis that all the 
absorption that we see in addition to the Galactic value is concentrated at the GRB host galaxy. 
The number of photons depends on the contrast with respect to the Galactic column density and on 
a combination of the intrinsic column density and redshift, defined by the effective column density $N_H(z=0)$.
The requested number of photons is very high but a few bursts observed within a few minutes from the 
GRB onset can meet the requirements. This technique is also promising for new large area, fast-slewing 
X--ray instruments (e.g. Campana et al. 2011b).

We applied this technique to the case of the XMM-Newton observation of GRB 090618. We successfully recovered
the GRB correct redshift with a $\sim 20\%$ ($90\%$ confidence level) with a factor of $\sim 3$ less counts 
than predicted to have a $10\%$ accuracy. However the improvement obtained leaving free the redshift parameter is 
only marginal ($8\%$ based on an F-test), calling for a larger number of photons.
In fact a low-confidence, small region consistent with $z=0$ is allowed in the $N_H(z)-z$ plane.
The inclusion of $\sim 6,000$ high-quality photons from the RGS instruments, leads to a 
better characterisation of the redshift and a much higher significance for the need of a host galaxy absorber 
(F-test chance improvement probability of $\sim 3\,\sigma$). 

We concluded that the most important parameter to derive the GRB redshift from X--ray data alone is the number of counts.
Therefore, a large effective area coupled to an energy band extending down to, at least, 0.3 keV
(and possibly 0.1--0.2 keV to deal with high redshift GRBs) would provide an optimal instrument for this kind of studies.
A very good spectral resolution would provide further improvements leading to the detection of single
absorption edges and opening the possibility to study the material composition.
Athena X-IFU is the best proposed instrument for this kind of studies envisaged for the near future (Jonker et al. 2013).

\section{Acknowledgments}
We thank S. Covino for useful conversations.
We thank ASI (I/004/11/1) and PRIN-MIUR (2009ERC3HT) for support.

\end{document}